\documentclass[twoside,twocolumn,10pt]{article}
\usepackage{extsizes}
\usepackage[super,sort&compress,comma]{natbib} 
\usepackage[version=3]{mhchem}
\usepackage[left=1.5cm, right=1.5cm, top=1.785cm, bottom=2.0cm]{geometry}
\usepackage{balance}
\usepackage{mathptmx}
\usepackage{sectsty}
\usepackage{graphicx} 
\usepackage{lastpage}
\usepackage[format=plain,justification=justified,singlelinecheck=false,font={stretch=1.125,small,sf},labelfont=bf,labelsep=space]{caption}
\usepackage{float}
\usepackage{fancyhdr}
\usepackage{fnpos}
\usepackage[english]{babel}
\addto{\captionsenglish}{%
  
}
\usepackage{array}
\usepackage{droidsans}
\usepackage{charter}
\usepackage[T1]{fontenc}
\usepackage[usenames,dvipsnames]{xcolor}
\usepackage{setspace}
\usepackage[compact]{titlesec}
\usepackage{hyperref}

\usepackage{epstopdf}

\definecolor{cream}{RGB}{222,217,201}
\usepackage{colortbl}
\usepackage{booktabs}

\begin{document}

\pagestyle{fancy}
\thispagestyle{plain}
\fancypagestyle{plain}{
\renewcommand{\headrulewidth}{0pt}
}

\makeFNbottom
\makeatletter
\renewcommand\LARGE{\@setfontsize\LARGE{15pt}{17}}
\renewcommand\Large{\@setfontsize\Large{12pt}{14}}
\renewcommand\large{\@setfontsize\large{10pt}{12}}
\renewcommand\footnotesize{\@setfontsize\footnotesize{7pt}{10}}
\makeatother

\renewcommand{\thefootnote}{\fnsymbol{footnote}}
\renewcommand\footnoterule{\vspace*{1pt}%
\color{cream}\hrule width 3.5in height 0.4pt \color{black}\vspace*{5pt}} 
\setcounter{secnumdepth}{5}

\makeatletter 
\renewcommand\@biblabel[1]{#1}            
\renewcommand\@makefntext[1]%
{\noindent\makebox[0pt][r]{\@thefnmark\,}#1}
\makeatother 
\renewcommand{\figurename}{\small{Fig.}~}
\sectionfont{\sffamily\Large}
\subsectionfont{\normalsize}
\subsubsectionfont{\bf}
\setstretch{1.125} 
\setlength{\skip\footins}{0.8cm}
\setlength{\footnotesep}{0.25cm}
\setlength{\jot}{10pt}
\titlespacing*{\section}{0pt}{4pt}{4pt}
\titlespacing*{\subsection}{0pt}{15pt}{1pt}

\fancyfoot{}
\fancyfoot[LO,RE]{\vspace{-7.1pt}\includegraphics[height=9pt]{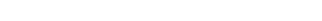}}
\fancyfoot[CO]{\vspace{-7.1pt}\hspace{13.2cm}\includegraphics{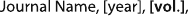}}
\fancyfoot[CE]{\vspace{-7.2pt}\hspace{-14.2cm}\includegraphics{head_foot/RF}}
\fancyfoot[RO]{\footnotesize{\sffamily{1--\pageref{LastPage} ~\textbar  \hspace{2pt}\thepage}}}
\fancyfoot[LE]{\footnotesize{\sffamily{\thepage~\textbar\hspace{3.45cm} 1--\pageref{LastPage}}}}
\fancyhead{}
\renewcommand{\headrulewidth}{0pt} 
\renewcommand{\footrulewidth}{0pt}
\setlength{\arrayrulewidth}{1pt}
\setlength{\columnsep}{6.5mm}
\setlength\bibsep{1pt}

\makeatletter 
\newlength{\figrulesep} 
\setlength{\figrulesep}{0.5\textfloatsep} 

\newcommand{\topfigrule}{\vspace*{-1pt}%
\noindent{\color{cream}\rule[-\figrulesep]{\columnwidth}{1.5pt}} }

\newcommand{\botfigrule}{\vspace*{-2pt}%
\noindent{\color{cream}\rule[\figrulesep]{\columnwidth}{1.5pt}} }

\newcommand{\dblfigrule}{\vspace*{-1pt}%
\noindent{\color{cream}\rule[-\figrulesep]{\textwidth}{1.5pt}} }

\makeatother

\twocolumn[
\begin{@twocolumnfalse}
{\includegraphics[height=30pt]{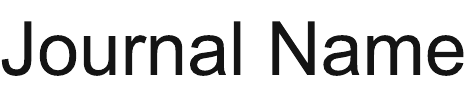}\hfill\raisebox{0pt}[0pt][0pt]{\includegraphics[height=55pt]{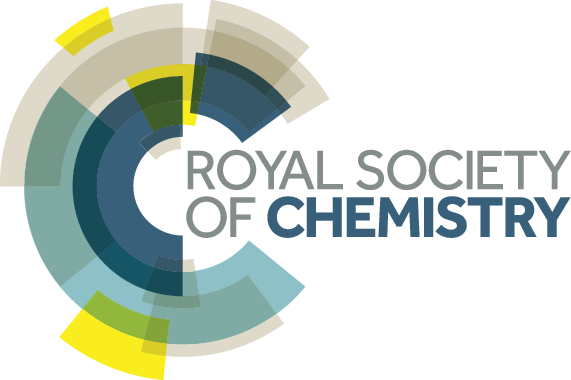}}\\[1ex]
\includegraphics[width=18.5cm]{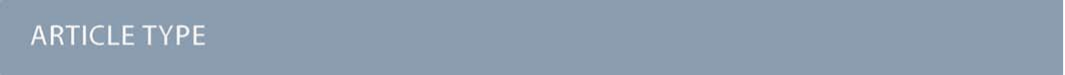}}\par
\vspace{1em}
\sffamily
\begin{tabular}{m{4.5cm} p{13.5cm} }

\includegraphics{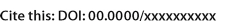} & \noindent\LARGE{\textbf{Crystallographic Challenges in Microscopy of Multi-domain Spinel Materials.}} \\
\vspace{0.3cm} & \vspace{0.3cm} \\

 & \noindent\large{Ninon Scherz,\textit{$^{a}$} Shashwat Anand,\textit{$^{b}$} Colin Ophus,\textit{$^{c}$} Tucker Holstun,\textit{$^{a,b}$}  Mary Scott,\textit{$^{a,d}$}  Tara P. Mishra,\textit{$^{d,\ddag}$} and Gerbrand Ceder \textit{$^{a,b,\ast}$}} \\

\includegraphics{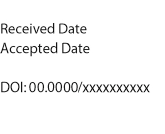} & \noindent\normalsize{Electron microscopy techniques are instrumental in the characterization of energy storage materials, with atomic resolution images providing the detailed structural features that are needed to understand their properties. Atomically resolved electron microscopy techniques have been routinely used to study the microstructure in high-performing Mn-based oxide cathodes, which often contain spinel-like ordering. Here, we evaluate STEM-HAADF imaging and subsequent Fourier filtering as tools for characterizing $\delta$-DRX spinel domains and their antiphase boundaries, which play a central role in the material’s electrochemical performance. Using electron microscopy simulations and recent theoretical insight into the structural makeup of $\delta$-DRX, we attempt to characterize the crystallographic spinel variants which occur in its multi-domain structure. We show that each domain interface, arising from pairings among eight distinct variants, can be categorized into one of four Fourier filtered profiles, one of which leaves the boundary undetectable in atomically resolved electron microscopy when viewed along the preferred [110] zone axis. Our results also suggest that the appearance of seemingly disordered or layered-like regions might actually arise from low energy domain boundaries which are slanted relative to the [110] viewing direction. Our findings highlight the need for careful interpretation of atomic-resolution micrographs of phase transitions, where local reordering drives transformations from higher to lower symmetry structures while maintaining lattice coherence.} 
\end{tabular}

\end{@twocolumnfalse} \vspace{0.6cm}

]

\renewcommand*\rmdefault{bch}\normalfont\upshape
\rmfamily
\section*{}
\vspace{-1cm}


\footnotetext{\textit{$^{a}$~Department of Materials Science and Engineering, University of California, Berkeley, CA,
USA.}}
\footnotetext{\textit{$^{b}$~Materials Sciences Division, Lawrence Berkeley National Laboratory, CA, USA. }}
\footnotetext{\textit{$^{c}$~Department of Materials Science and Engineering, Stanford University, Stanford, CA,
USA. }}
\footnotetext{\textit{$^{d}$~National Center for Electron Microscopy, Molecular Foundry, Lawrence Berkeley National
Laboratory, Berkeley, CA, USA. }}


\footnotetext{$\ddag$~email: tpmishra@lbl.gov}
\footnotetext{$\ast$~email: gceder@berkeley.edu}


\section{Introduction}
Electron microscopy is a pillar of modern materials characterization, offering atomic-resolution imaging of structure, unit-cell parameters, local symmetry, and defects \cite{gong2025energy_storage_materials_EM, smith2020atomic, yoon2024decoding, allen2012chemical}. Yet, its capacity to fully resolve three-dimensional crystallographic information remains fundamentally constrained by the problem of projection since the real-space micrographs are two-dimensional representations of three-dimensional lattices \cite{arslan2012_3D_atoms, bals2014seeing}. Consequently, complex three-dimensional arrangements can be challenging to interpret when viewed through two-dimensional projections.

Conventional methods such as High-Angle Annular Dark Field Scanning Transmission Electron Microscopy (STEM-HAADF) remain among the most accessible and widely implemented techniques in materials characterization \cite{spurgeon2021towards, cross2015materials}. Clarifying what STEM-HAADF imaging and its post-processing approaches can and cannot reliably reveal is therefore essential, especially in the study of structurally complex materials where macroscopic properties are governed by interfaces and small-scale features\cite{neumann1996crystallography, mannhar2012interface}. Understanding these limits helps bridge the gap between what is practically observable and what is theoretically understood, enabling a more informed interpretation of structure–property relationships at the atomic scale \cite{lee2025unravelling_CRYO_EM}.

Li-ion cathodes with high Mn content are of particular interest to electron microscopy \cite{
Raphael2020_DRX_Review_EES,
Lee2014_Science_DRX,
Lun2020_Chem_Mn_DRX,
chiang2000electrochemically,
jang1999_JECS_Ortho_to_Spinel,
YMChiang1999_ECSSLett_Origin_of_Transformation,
Pralong2016_NatMater_Li4Mn2O5,
Chongmin_Raphael_2022_AEM_delta}, as the high mobility of Mn \cite{holstun2025accelerating_Tucker} and the strong driving force to form spinel at \ce{LiMn2O4} composition \cite{mishra1999structural_grndstate, Paulsen1999_ChemMaterLiMnO_PD_grndstate} can lead to the emergence of small-scale structural features upon long-term cycling \cite{gummow1993lithium, reed2001layered, lee2014structural, chiang2000electrochemically, jang1999electrochemical, 2024insitu, wang1999origin, Chongmin_Raphael_2022_AEM_delta, Paulsen1999_ChemMaterLiMnO_PD_grndstate}. These small-scale features can be important for their influence on the voltage profile and Li-diffusion kinetics. More recently, Mn-rich oxides have gained renewed interest due to their high safety, low toxicity, and the high natural abundance of Mn.

The case study of this work, $\delta$-DRX, is an example of such Mn-rich oxides being actively studied using advanced electron microscopy \cite{holstun2025accelerating_Tucker,hau2024earth-abundant}. The unique structure of $\delta$-DRX emerges during the electrochemical cycling or chemical delithiation of Mn-rich disordered rocksalt (DRX) (e.g. \ce{Li_{1.05}Mn_{0.85}Ti_{0.1}O_2}), a cation-disordered structure with Fm$\bar{3}$m symmetry in which Li and transition metals share the octahedral sites\cite{holstun2025accelerating_Tucker,hau2024earth-abundant, li2024structural, li2025impact, Guoying_Raphaele_2023_AEM_delta}. The transformation rearranges the transition metal atoms on the cubic lattice to nucleate spinel order\cite{hau2024earth-abundant}. The transformed $\delta$-DRX material has a multi-domain nanostructure, where each ordered spinel domain is separated from its neighbors by thin boundaries\cite{holstun2025accelerating_Tucker}. This unique nanostructure of the $\delta$-DRX is helpful in suppressing the two-phase cubic-to-tetragonal phase transformation which limits the usable capacity of well-ordered spinel\cite{holstun2025accelerating_Tucker,anand2025origin} and in achieving better rate capability than DRX \cite{hau2024earth-abundant,holstun2025accelerating_Tucker}. Given the small extent of remnant disorder within the domains of $\delta$-DRX as evidenced by diffraction,\cite{hau2024earth-abundant, holstun2025accelerating_Tucker} a comprehensive characterization of the type of domain boundaries, domain morphology and domain size are crucial for understanding its ionic transport and suppression of the detrimental two-phase reaction.

Recently, the crystallography of multi-domain spinel nanostructures has been clarified using group theory analysis \cite{anand2025origin}. As the quotient of the rocksalt space group (\textit{Fm}$\overline{3}$m) and spinel space group (\textit{Fd}$\overline{3}$m) has cardinality of 8, eight distinct variants of spinel ordering can form on the parent rocksalt lattice. These variants are related by rotational and translational symmetry elements of the rocksalt space group that are not symmetry operations of the spinel space group.

This symmetry breaking manifests microstructurally in the formation of spinel domains in $\delta$-DRX. In the spinel space group, the octahedral sublattice partitions into two distinct Wyckoff sites, conventionally denoted as 16c (vacant) and
16d (Mn-occupied), while Li resides on tetrahedral sites. While all octahedral sites are symmetry-equivalent in the parent disordered rocksalt phase, the loss of symmetry generates an additional degree of freedom for the arrangement of transition metals. The specific spatial configuration of these $16c/16d$ sublattices defines the eight "spinel variants." The multi-domain structure of $\delta$-DRX arises as different regions of the crystal lattice independently adopt one of these permitted variants. These distinct arrangements of spinel occupancy are what render the domains distinguishable in atomic-resolution STEM-HAADF when viewed along the [110] zone axis.

In this work, we evaluate the extent to which atomically resolved STEM–HAADF can detect and characterize $\delta$-DRX antiphase boundaries, through analysis of STEM–HAADF simulations of modeled interfaces between spinel variants viewed along the most experimentally relevant [110] zone axis. We use Fourier filtering as a tool to aid STEM–HAADF image interpretation, as it isolates spinel spatial frequencies and facilitates the analysis of ambiguous projected contrast. We show that the appearance of seemingly disordered or layered regions in STEM–HAADF may arise from low-energy antiphase boundaries (\{100\}) \cite{anand2025origin} that are slanted relative to the [110] viewing direction. Our results further demonstrate that antiphase boundaries between some pairs of spinel variants are undetectable in atomically resolved STEM–HAADF when viewed along the [110] zone axis. Taken together, these findings reveal subtle crystallographic challenges inherent to the microscopy of multi-domain systems and are broadly relevant to other materials that require precise characterization of domain boundaries on coherent lattices \cite{wang2018designing, xu2023topotactically, chen2025spinel}.

\section{Results}
\subsection{Crystallographic variants formed during symmetry reducing phase transformation}

\begin{figure}[!t]
\centering
    \includegraphics[width=\columnwidth]{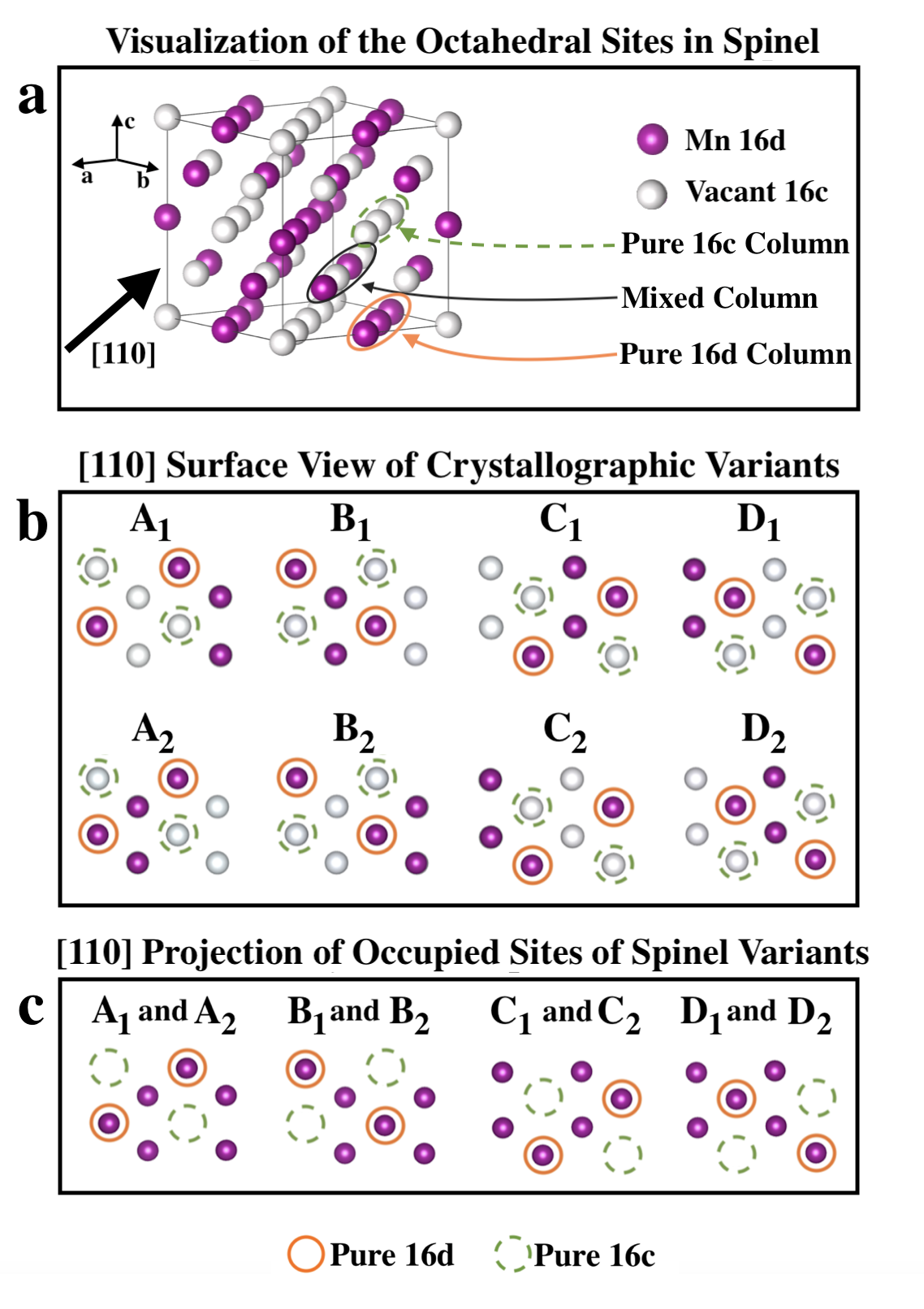}
    \caption{a) 3D visualization of the Mn-occupied $16d$, and the vacant $16c$ sites of the spinel unit cell. Arrows point to atomic columns forming in the [110] zone axis: the dotted green arrow points to a pure $16c$ (vacant) column, the black arrow points to a column of mixed occupancy, and the orange arrow points to a pure $16d$ (occupied) column.
    b) [110] view of the atomic sites on the surface of the eight spinel variants (named $A_i$, $B_i$, $C_{i}$ and $D_i$, where $i=$1 and 2) on the parent rock-salt lattice. The sites from the $16c$ and $16d$ sub-lattices corresponding to each variant of spinel are shown with white and purple respectively. The sites beneath the top surface layer are hidden by the surface sites. The encircled sites represent the column of sites beneath the surface (extending into the paper) that either belong entirely to the $16c$ (dashed circles) or the $16d$ (solid circles) sub-lattice; and c) The [110] view of the variants shown in a) after removing all the vacancy sites. Mn sites hidden beneath the surface vacancy sites in (a) are visible in (b).}
    \label{fgr:Structure}
\end{figure}

Figure \ref{fgr:Structure}a presents a three-dimensional representation of a spinel unit cell in which only the octahedral sublattice is shown. The $16d$ and $16c$ Wyckoff sites of a fully ordered spinel are shown with purple and white spheres, respectively. Atomic columns that form along the [110] zone axis are indicated by arrows. The dotted green arrow marks a column composed solely of vacant $16c$ sites, the black arrow marks a column with mixed $16c$/$16d$ occupancy, and the orange arrow marks a column consisting purely of occupied $16d$ sites. Figure \ref{fgr:Structure}b shows a simplified visualization of the eight crystallographically distinct spinel variants (named $A_{i}$, $B_{i}$, $C_{i}$, and $D_i$, where $i=1, 2$) as viewed from the [110] direction. These variants are formed during the transformation from the high-symmetry $Fm\bar{3}m$ space group of rocksalt to the lower-symmetry $Fd\bar{3}m$ space group of spinel. We adopted this nomenclature to clarify the symmetrical relationships between the spinel variants in 2D STEM–HAADF projections, which is central to our discussion of the modeled interfaces. We note here that our previous work \cite{anand2025origin} adopted a different nomenclature which was more relevant to understanding 3-dimensional symmetry relationships between spinel variant. A mapping between the two nomenclatures is provided in Supplementary Table S1. Here, we classify the spinel variants into four pairs: $A_{i}$, $B_{i}$, $C_{i}$ and $D_i$ (where $i=1$ and $i=2$). For each pair, $i=1$ and $i=2$ are related to each other by a twofold rotation axis along the ``pure" $16c$ or $16d$ site columns. Furthermore, the $A_{i}$ ($C_i$) variants are related to the $B_i$ ($D_i$) variants through a complete exchange in occupancies of the vacant and occupied sub-lattices (i.e swapping between the 16c and 16d sites). The $A$/$B$ family differ from the $C$/$D$ family by a lattice translation of the spinel ordering relative to the rocksalt framework, which interchanges mixed and pure ($16c$/$16d$)
columns in a distinct way. All variants can be related to each other through translational operations on the cubic lattice.

Since the definition of the octahedral occupancy (Mn or Li/vacancy) alone is sufficient to differentiate all eight crystallographic variants, only octahedral sites are shown; accordingly Figures \ref{fgr:Structure}b and \ref{fgr:Structure}c simplify the spinel structure of Figure \ref{fgr:Structure}a, using a reduced cell in the [110] projection. In Figures \ref{fgr:Structure}b and \ref{fgr:Structure}c, all visible sites lie on the surface layer, and the octahedral sites from the layers beneath are hidden by the surface sites. Although the $16c$ sites are only occupied upon full lithiation in the actual structure, they are included in Figures \ref{fgr:Structure}a and \ref{fgr:Structure}b as spatial placeholders to clarify the relative depths of the $16d$ sub-lattice in projection. Li, which resides on the tetrahedral $8a$ sites, is omitted here because the symmetry distinctions between the eight spinel variants arise from ordering on the octahedral sublattice. Moreover, Li is largely invisible in STEM-HAADF imaging due to its low atomic number, so excluding it does not affect how the cation ordering is interpreted in projection. This representation emphasizes the subtle three-dimensional offsets that distinguish the eight variants when viewed along the [110] zone axis.

In cases where the surface site and the sites beneath it are of the same type ($16c$ or $16d$), the sites are encircled (dashed or solid circle) and are referred to as ``pure" columns to indicate the unmixed occupancy of atomic sites along the [110] zone axis. We draw attention to these columns of unmixed occupancy because they are crucial for creating the  contrast necessary for distinguishing spinel ordering on the rocksalt lattice from the cation disordered structure in STEM-HAADF projections. In contrast to the spinel ordered structure, all atomic columns of the cation disordered structure (DRX) have a mixed Mn and vacancy occupancy (50$\%$) along the [110] depth of the view. A comparison of the [110] view with other low-index zone axes such as [100] and [111] in Supplementary Figure S1 shows that the $16c$ and $16d$ Wyckoff sites overlap along those directions. This illustrates why the [110] axis serves as the primary orientation for imaging the $\delta$ phase and resolving the different domains formed during the symmetry-reducing phase transformation.

\begin{figure*}[!t]
    \centering
    \includegraphics[width=0.9\textwidth]{Figures/Figure_2.png}
    \caption{ a) Experimental STEM-HAADF image of $\delta$-DRX. 
b) FFT of a), with spinel frequencies indexed at $(\bar{1}1\bar{1})$ and $(\bar{1}11)$ circled in cyan and red, and parent lattice frequencies indexed at $(\bar{2}2\bar{2})$ and $(\bar{2}22)$ circled in white. 
c) Fourier-filtered HAADF image using the $(\bar{1}1\bar{1})$ type frequency (red) and a DRX-like frequency indexed at the $(\bar{2}2\bar{2})$ parent-lattice position (white). 
d) Fourier-filtered HAADF image using the $(\bar{1}11)$ type frequency (cyan) and a DRX-like frequency indexed at the $(\bar{2}22)$ parent-lattice position (white). 
Green arrows indicate fringe offsets that reveal discontinuities. The dashed symbols correspond to four interface boundary types: the circle highlights a region with no discontinuity in either frequency; the triangle marks a region discontinuous in both; the pentagon marks a region discontinuous only in the $(\bar{1}1\bar{1})$ type frequency; and the square marks a region continuous in the $(\bar{1}1\bar{1})$ type frequency but discontinuous in the $(\bar{1}11)$ type frequency.
}
    \label{fig:experimental}
\end{figure*}

To understand how the spinel variants appear in experimental STEM-HAADF imaging, Figure~\ref{fgr:Structure}c shows the same [110] projection as Figure~\ref{fgr:Structure}b but without marking the white vacancy sites. In this view, the previously hidden subsurface cation sites are visible, and the ``pure'' $16c$ columns project as empty. In the [110] projection, the eight variants collapse into four projected motifs, each shared by a pair within a family: ($A_1$,$A_2$), ($B_1$,$B_2$), ($C_1$,$C_2$), and ($D_1$,$D_2$), differing only by the relative depth of atoms along the beam direction. These depth differences do not affect HAADF contrast in a simple projection, making the $i=1$ and $i=2$ variants within each pair indistinguishable in [110]. 
This projection degeneracy, and the absence of alternative low-index zone axes that recover this information, is discussed in Supplementary Note 1.

Figure \ref{fig:experimental}a presents an experimental atomic-resolution STEM-HAADF image of $\delta$-DRX. Several regions are highlighted by dashed outlines to indicate distinct local orderings. Figure \ref{fig:experimental}b shows the Fourier transform of the STEM-HAADF image. In the Fourier transform, each point represents a spatial frequency contained in the original image, where spatial frequency is defined as the inverse of the real-space spacing of a repeating feature. The top (red) and bottom (cyan) circles mark frequency spots associated with spinel-like frequencies, while white circles denote those corresponding to DRX-like frequencies. To facilitate comparison with crystallographic conventions, spatial frequencies observed in the FFTs of experimental and simulated STEM-HAADF images are indexed using diffraction-style notation. Although these FFTs arise from projected real-space intensity modulations rather than electron diffraction, the indexed frequencies correspond directly to periodicities associated with specific lattice planes in projection. In particular, spinel ordering gives rise to superlattice frequencies indexed at $(111)$ positions, while frequencies appearing at $(222)$ positions correspond to the parent-lattice periodicity characteristic of DRX. Throughout this work, such indexing is used to identify the spatial periodicities present in projection.

Figures~\ref{fig:experimental}c and~\ref{fig:experimental}d respectively show the STEM-HAADF image filtered for the $(\bar{1}1\bar{1})$ type and $(\bar{1}11)$ type spinel spatial frequencies, and for the $(\bar{2}2\bar{2})$ and $(\bar{2}22)$ DRX-like frequencies. The white fringes correspond to the DRX-like spatial frequency component that appears at twice the reciprocal-space radius of the spinel peaks. Because the spinel phase forms within the parent DRX framework, the underlying DRX periodicity is retained throughout the structure. In the filtered images shown here, DRX-like frequencies are intentionally suppressed in regions of strong spinel contrast to enable clear identification of where spinel ordering is continuous or disrupted, while allowing DRX-like fringes to persist only where spinel contrast is discontinuous or absent. Green arrows highlight offsets in the filtered spinel fringes, marking antiphase boundaries where the local cation ordering changes between spinel variants. In these boundary regions, the DRX-like frequency often appears between spinel domains.

White dashed shapes in Figures \ref{fig:experimental}a, \ref{fig:experimental}c, and \ref{fig:experimental}d indicate different cases of fringe continuity: the circle marks a region where both $(\bar{1}1\bar{1})$ type and $(\bar{1}11)$ type spinel fringes are continuous; the triangle marks a region where both are discontinuous; the pentagon marks a region where only the $(\bar{1}1\bar{1})$ type fringes are discontinuous; and the square marks a region where only the $(\bar{1}11)$ type fringes are discontinuous. The purpose of this experimental example is not to establish the statistical prevalence of particular interface classes within a given specimen, but rather to illustrate the diversity of discontinuity profiles that may arise in experimentally observed $\delta$-DRX microstructures. The crystallographic variant relationships giving rise to these discontinuity profiles cannot be uniquely inferred from the experimental image alone, motivating the systematic analysis of all crystallographically possible variant relationships presented in the following sections.

\subsection{Fourier-Filtered STEM Analysis of Discontinuities at Simulated Spinel Interfaces}

\begin{figure*}[!t]
    \centering
    \includegraphics[width=0.8\textwidth, keepaspectratio]{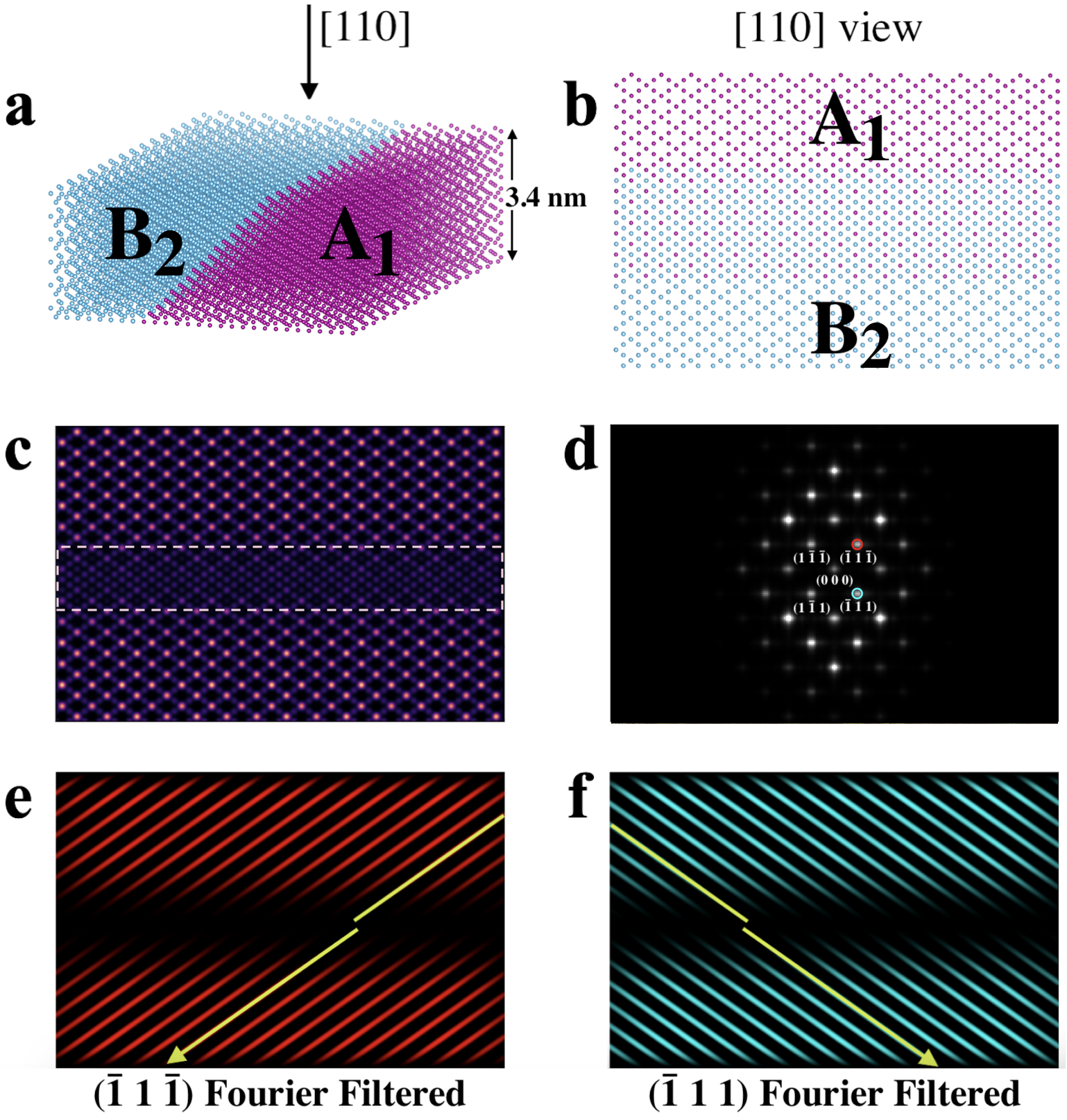}
    \caption{
a) Slanted interface boundary construction between the $A_{1}$ and $B_{2}$ spinel variants.
b) [110] visualization of the constructed interfacial structure.
c) Simulated [110] STEM-HAADF-like contrast of the interfacial structure. The dashed box highlights the region projecting the slanted {100} interface.
d) Adjusted FFT of c), showing two isolated spinel-related spatial frequencies indexed at $(\bar{1}1\bar{1})$ and $(\bar{1}11)$, circled in red and cyan, respectively.
e) Fourier-filtered HAADF image for the $(\bar{1}1\bar{1})$-type frequency.
f) Fourier-filtered HAADF image for the $(\bar{1}11)$-type frequency.
}
    \label{fig:100_Interface_Example}
\end{figure*}

To simulate realistic spinel antiphase boundaries in $\delta$-DRX using STEM-HAADF, we construct \{100\} interfaces between spinel variants. These interfaces were previously calculated to have the lowest formation energies for most variant pairings~\cite{anand2025origin}, making them the most physically relevant boundaries to model. When viewed along the [110] direction, a \{100\} boundary appears as a slanted interface tilted by $45^\circ$ within the projected slab. Figure~\ref{fig:100_Interface_Example}a illustrates an example of such a boundary between the $A_{1}$ and $B_{2}$ variants. As in Figure~\ref{fgr:Structure}c, only the occupied octahedral sublattices are displayed in Figures~\ref{fig:100_Interface_Example}a and~\ref{fig:100_Interface_Example}b to highlight the cation ordering. To clearly visualize the \{100\} interface, Mn atoms belonging to the two domains are shown in different colors:  $A_{1}$ Mn in purple, and $B_{2}$ Mn in blue.

Figure \ref{fig:100_Interface_Example}b shows the top view of the same slab as in panel a, looking down the [110] direction. For this pair of spinel variants ($A_{1}$-$B_{2}$), the top view has 3 distinct regions. The top and bottom regions correspond to the projection of the regions where only one spinel domain exists along the [110] axis. The middle region, however, represents the slanted \{100\} interface between the domains, and atoms from both domains are visible in the projection.

The structure in Figure \ref{fig:100_Interface_Example}b is used in atomic resolution STEM-HAADF simulation in Figure \ref{fig:100_Interface_Example}c. The ordered regions at the top and bottom, and the interfacial region in the center is clearly distinguishable. The high intensity (bright) sites in the ordered regions correspond to the pure $16d$ columns in Figure \ref{fgr:Structure}b. Between these high-intensity sites, are much lower intensity sites which correspond to the columns with mixed Mn/Vac occupancy in Figure \ref{fgr:Structure}b. The columns corresponding to the pure $16c$ sites are not visible in the ordered spinel regions, since they are vacant. The middle interface region projects all octahedral sites ($16d$ and $16c$), but at a much lower intensity because all columns in this region have a mixed Mn/Vac occupancy along the depth of view. 

The simulated STEM-HAADF image in Figure~\ref{fig:100_Interface_Example}c is Fourier transformed to obtain the frequency space representation in Figure~\ref{fig:100_Interface_Example}d. As in the experimental transform, features with large real-space spacings appear near the center, while features with small spacings appear at higher spatial frequencies farther from the center. By placing masks around the specific frequency peaks associated with spinel-like ordering, we isolate those components, and applying the inverse Fourier transform yields real-space images that contain only the selected structural modulations.

The FFT allows us to isolate the two distinct spinel lattice periodicities. Figures~\ref{fig:100_Interface_Example}e and~\ref{fig:100_Interface_Example}f show the image in Figure~\ref{fig:100_Interface_Example}c filtered for the $(\bar{1}1\bar{1})$ (red) and $(\bar{1}11)$ (cyan) spatial frequencies, respectively. Green arrows mark discontinuities in each filtered image, indicating interruptions in the corresponding structural frequency. Similar discontinuities appear in the experimental data (Figures~\ref{fig:experimental}c and~\ref{fig:experimental}d). For this specific interface combination ($A_{1}$--$B_{2}$), both filtered components exhibit two discontinuities within the interface region. 
We note that the apparent width of the simulated \{100\} interface in Figure \ref{fig:100_Interface_Example}b--c arises from projection geometry rather than from an intrinsically broad structural transition. A \{100\} boundary viewed along [110] intersects the slab at 45°, so its projected width increases with the simulated sample thickness.

The steps detailed in Figure \ref{fig:100_Interface_Example} were performed on simulated \{100\} interfaces between every spinel variant combination. These include simulating a STEM-HAADF micrograph from the structure constructed with different spinel variants interfaces along the \{100\} plane, performing a FFT on the STEM-HAADF micrograph, masking spinel frequencies, followed by inverse Fourier transformation to isolate the $(\bar{1}1\bar{1})$ and $(\bar{1}11)$ type fringes in filtered results. Supplementary Figures S2–S29 include every STEM–HAADF simulation and corresponding Fourier filtered results.

\begin{figure*}[!t]
    \centering
    \includegraphics[width=0.9\textwidth]{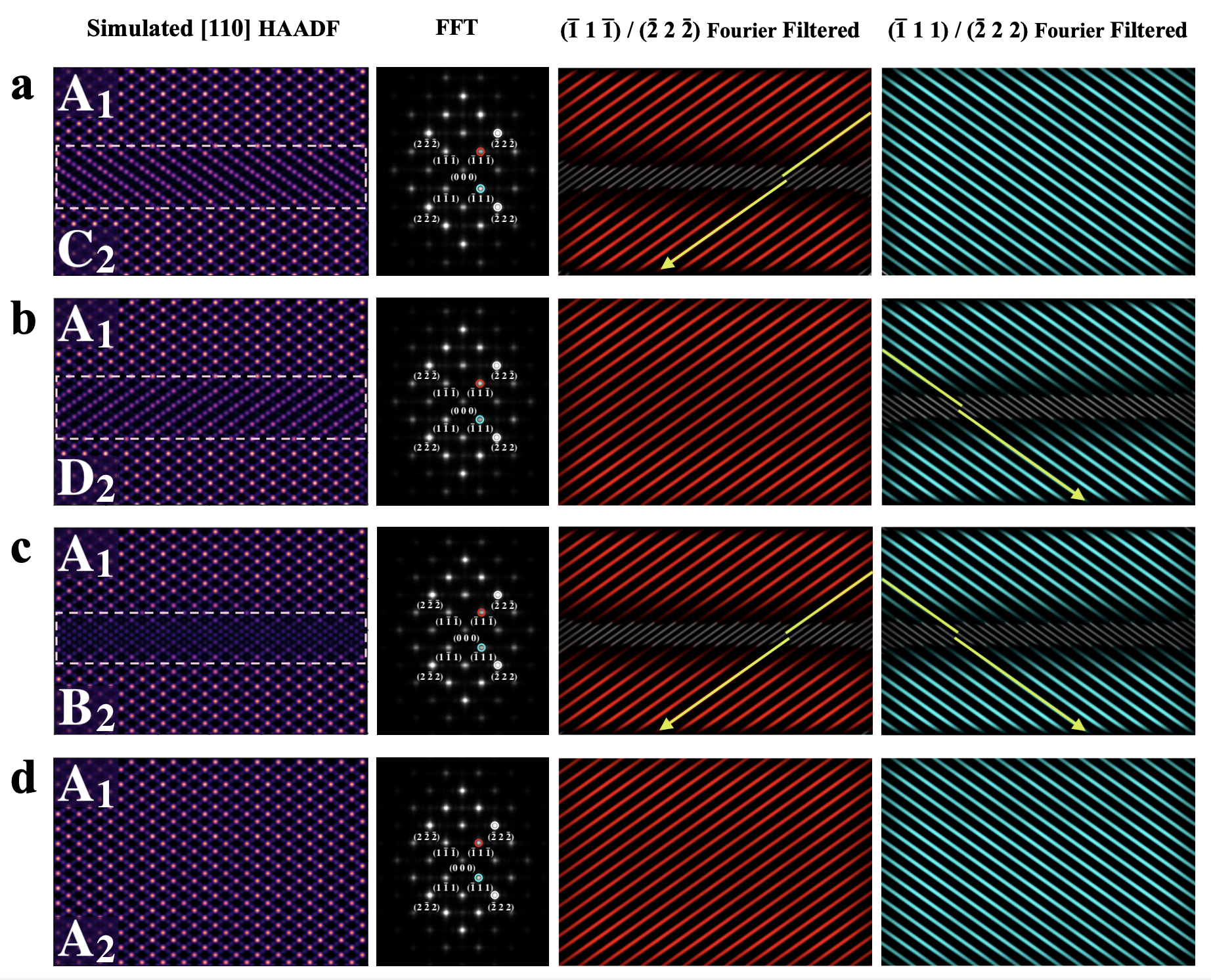}
    \caption{Four distinct cases of variant interfaces. $(\bar{1}1\bar{1})$ type and $(\bar{1}11)$ type fringes are respectively shown by isolating the top and bottom spinel frequencies identified in the FFT. The spots circled in white in the FFT correspond to the parent-lattice $(\bar{2}2\bar{2})$ and $(\bar{2}22)$ frequencies. The dashed boxes highlight the region projecting the slanted \{100\} interface in the STEM-HAADF simulation. White fringes in the filtered images indicate the presence of a DRX-like frequency indexed at the parent-lattice position.
}
    \label{fig:cases}
\end{figure*}

In Figure \ref{fig:cases}, we present four different cases. In Figure \ref{fig:cases}a, the interface between $A_{1}$ and $C_{2}$ forms a discontinuity only in the $(\bar{1}1\bar{1})$ type fringes; in Figure \ref{fig:cases}b, the $A_{1}$ and $D_{2}$ interface forms a discontinuity only in the $(\bar{1}11)$ type fringes; in Figure \ref{fig:cases}c, the $A_{1}$ and $B_{2}$ interface forms a discontinuity in both $(\bar{1}1\bar{1})$ and $(\bar{1}11)$ type fringes; and in Figure \ref{fig:cases}d, the $A_{1}$ and $A_{2}$ interface filtered images remain continuous for both $(\bar{1}1\bar{1})$ and $(\bar{1}11)$ type fringes. These four cases make up the four profiles within which all interface boundaries fall.
\begin{table*}[!h]
    \centering
    \caption{Discontinuity relationships between all spinel variant pairs for $(\bar{1}1\bar{1})$ type and $(\bar{1}11)$ type fringes. The variant pairs in grey cells do not have a visible discontinuity at the boundary in STEM-HAADF simulations.}
    \label{tab:discontinuity}
\begin{tabular}{|c|c||c|c|||c|||}
    \toprule
    \multicolumn{2}{|c|}{Spinel Interface} & \multicolumn{2}{||c|||}{Fourier Filter Discontinuity} & \textbf{Domain Boundary} \\
    \cmidrule(lr){1-4}
    Variant 1 
      & Variant 2
      & \textcolor{red}{$(\bar{1}1\bar{1})$ Type Fringes}
      & \textcolor{cyan}{$(\bar{1}11)$ Type Fringes}
      & \textbf{detection} \\
    \midrule
\cellcolor{gray!15}A$_{1}$ & \cellcolor{gray!15}A$_{2}$ & \cellcolor{gray!5}No & \cellcolor{gray!5}No & \cellcolor{gray!5}No \\
\cellcolor{gray!15}B$_{1}$ & \cellcolor{gray!15}B$_{2}$ & \cellcolor{gray!5}No & \cellcolor{gray!5}No & \cellcolor{gray!5}No \\
\cellcolor{gray!15}C$_{1}$ & \cellcolor{gray!15}C$_{2}$ & \cellcolor{gray!5}No & \cellcolor{gray!5}No & \cellcolor{gray!5}No \\
\cellcolor{gray!15}D$_{1}$ & \cellcolor{gray!15}D$_{2}$ & \cellcolor{gray!5}No & \cellcolor{gray!5}No & \cellcolor{gray!5}No \\
A$_{1}$ & B$_{1}$ & \cellcolor{green!25}Yes & \cellcolor{green!25}Yes & \cellcolor{green!25}Yes \\
A$_{1}$ & B$_{2}$ & \cellcolor{green!25}Yes & \cellcolor{green!25}Yes & \cellcolor{green!25}Yes \\
A$_{2}$ & B$_{1}$ & \cellcolor{green!25}Yes & \cellcolor{green!25}Yes & \cellcolor{green!25}Yes \\
A$_{2}$ & B$_{2}$ & \cellcolor{green!25}Yes & \cellcolor{green!25}Yes & \cellcolor{green!25}Yes \\
C$_{1}$ & D$_{1}$ & \cellcolor{green!25}Yes & \cellcolor{green!25}Yes & \cellcolor{green!25}Yes \\
C$_{1}$ & D$_{2}$ & \cellcolor{green!25}Yes & \cellcolor{green!25}Yes & \cellcolor{green!25}Yes \\
C$_{2}$ & D$_{1}$ & \cellcolor{green!25}Yes & \cellcolor{green!25}Yes & \cellcolor{green!25}Yes \\
C$_{2}$ & D$_{2}$ & \cellcolor{green!25}Yes & \cellcolor{green!25}Yes & \cellcolor{green!25}Yes \\
A$_{1}$ & C$_{1}$ & \cellcolor{green!25}Yes & \cellcolor{gray!5}No & \cellcolor{green!25}Yes \\
A$_{1}$ & C$_{2}$ & \cellcolor{green!25}Yes & \cellcolor{gray!5}No & \cellcolor{green!25}Yes \\
A$_{2}$ & C$_{1}$ & \cellcolor{green!25}Yes & \cellcolor{gray!5}No & \cellcolor{green!25}Yes \\
A$_{2}$ & C$_{2}$ & \cellcolor{green!25}Yes & \cellcolor{gray!5}No & \cellcolor{green!25}Yes \\
B$_{1}$ & D$_{1}$ & \cellcolor{green!25}Yes & \cellcolor{gray!5}No & \cellcolor{green!25}Yes \\
B$_{1}$ & D$_{2}$ & \cellcolor{green!25}Yes & \cellcolor{gray!5}No & \cellcolor{green!25}Yes \\
B$_{2}$ & D$_{1}$ & \cellcolor{green!25}Yes & \cellcolor{gray!5}No & \cellcolor{green!25}Yes \\
B$_{2}$ & D$_{2}$ & \cellcolor{green!25}Yes & \cellcolor{gray!5}No & \cellcolor{green!25}Yes \\
A$_{1}$ & D$_{1}$ & \cellcolor{gray!5}No & \cellcolor{green!25}Yes & \cellcolor{green!25}Yes \\
A$_{1}$ & D$_{2}$ & \cellcolor{gray!5}No & \cellcolor{green!25}Yes & \cellcolor{green!25}Yes \\
A$_{2}$ & D$_{1}$ & \cellcolor{gray!5}No & \cellcolor{green!25}Yes & \cellcolor{green!25}Yes \\
A$_{2}$ & D$_{2}$ & \cellcolor{gray!5}No & \cellcolor{green!25}Yes & \cellcolor{green!25}Yes \\
B$_{1}$ & C$_{1}$ & \cellcolor{gray!5}No & \cellcolor{green!25}Yes & \cellcolor{green!25}Yes \\
B$_{1}$ & C$_{2}$ & \cellcolor{gray!5}No & \cellcolor{green!25}Yes & \cellcolor{green!25}Yes \\
B$_{2}$ & C$_{1}$ & \cellcolor{gray!5}No & \cellcolor{green!25}Yes & \cellcolor{green!25}Yes \\
B$_{2}$ & C$_{2}$ & \cellcolor{gray!5}No & \cellcolor{green!25}Yes & \cellcolor{green!25}Yes \\
    \bottomrule
\end{tabular}
\end{table*}

In Figure~\ref{fig:100_Interface_Example}c--f, the STEM–HAADF image is filtered only for the two spinel spatial frequencies: one filtered image isolates the $(\bar{1}1\bar{1})$ type component, and the other isolates the $(\bar{1}11)$ type component. In Figure~\ref{fig:cases}, we apply the same two spinel filters, but additionally select the DRX-like spatial frequency, indexed at the parent-lattice $(\bar{2}2\bar{2})$ and $(\bar{2}22)$ positions, corresponding to the higher-frequency lattice fringes. This DRX frequency component is present throughout the image and, if not suppressed, overlaps with the spinel fringes. To facilitate the visualization of the ordered spinel domains, we suppress the DRX contribution within regions that exhibit clear spinel ordering, allowing the DRX-like fringes to remain visible only at the interfaces. Supplementary Figure S30 compares Fourier-filtered images generated with and without this selective suppression. In both cases, the discontinuities in the spinel-ordering frequencies remain clearly identifiable and lead to the same interface classification.
Figures~\ref{fig:cases}a--\ref{fig:cases}c show that the selected DRX-like frequency appears as white fringes in the slanted \{100\} interface region. In Figure~\ref{fig:cases}c, corresponding to the $A_{1}$-$B_{2}$ boundary shown earlier in Figure \ref{fig:100_Interface_Example}, the DRX-like fringes are visible in both the $(\bar{1}1\bar{1})$ and $(\bar{1}11)$ filtered images.

The STEM-HAADF simulations of Figures~\ref{fig:cases}a--\ref{fig:cases}d show top and bottom regions displaying single spinel variant orderings along the [110] direction. As in Figure~\ref{fig:100_Interface_Example}, variations in HAADF intensity arise from the projection of pure and mixed $16d/16c$ columns across the interface. 

 The interface region, boxed with a dashed line, shows a visibly different ordering in cases \ref{fig:cases}a, \ref{fig:cases}b and \ref{fig:cases}c. In Figures \ref{fig:cases}a and \ref{fig:cases}b, the boxed interface projection resembles that of layered structure, with layers composed of atomic columns of varying intensity. These layered atomic columns result from mixed $16d$-$16c$ columns, as they are dimmer than the highest intensity atomic columns of the pure $16d$ columns of the ordered regions. Some extremely faint atomic columns do appear between the brighter layers, showing some mostly vacant mixed column, with very few occupied $16d$ sites.

As discussed for Figure \ref{fig:100_Interface_Example}c, the interface region of \ref{fig:cases}c resembles a disordered structure, with atomic columns projecting mixed $16d$-$16c$ columns. In Figures~\ref{fig:cases}a--c, the overlap region of two ordered spinel domains across a coherent interface produces projected images that do not resemble spinel ordering. 
In Figure \ref{fig:cases}d, the two ordered spinel regions of $A_{1}$ and $A_{2}$ create an interface which does not interrupt the projected ordering, and where the spinel order is the same throughout the whole image and across the interface.

Table~\ref{tab:discontinuity} summarizes all simulated results, including the Fourier filter discontinuity profile of each spinel variant pair, as well as the presence or absence of interface detection in simulated STEM–HAADF images. The complete set of STEM–HAADF simulations and their corresponding $(\bar{1}1\bar{1})$ and $(\bar{1}11)$ Fourier filtered results for all simulated spinel variant pairs is provided in Supplementary Figures S2–S29. Four pairs of variants create interfaces that are undetectable from the [110] direction: $A_{1}$ and $A_{2}$, $B_{1}$ and $B_{2}$, $C_{1}$ and $C_{2}$, $D_{1}$ and $D_{2}$. All other spinel variant pairs are detectable: eight pairs are discontinuous in both $(\bar{1}1\bar{1})$ and $(\bar{1}11)$ type fringes, eight pairs are discontinuous only in $(\bar{1}1\bar{1})$ type fringes, and eight pairs are discontinuous only in $(\bar{1}11)$ type fringes. Among the eight pairs sharing the same Fourier discontinuity profile, one variant pair cannot be distinguished from another when viewed along the [110] direction.

\section{Discussion}

The simulated images shown throughout this work are projected electrostatic potentials generated from the structural models and used to produce HAADF-like contrast emphasizing the projected cation ordering. Since the objective of this study is to determine how crystallographic variant relationships manifest in projection rather than to quantitatively reproduce experimental image intensities, this approach provides a computationally efficient means of identifying the projected ordering motifs and interface discontinuities associated with different interface geometries. This work defines the limits of single-zone-axis STEM-HAADF imaging during symmetry-reduction transformations, as is the case when spinel ordering forms from a rocksalt. While multiple zone axes can sometimes provide three-dimensional structural data \cite{gong2018three}, characterizing spinel domain formation in $\delta$-DRX is effectively restricted to the [110] zone axis. As shown in Figure S1, other low-index axes (e.g., [100] and [111]) result in overlapping projections of spinel $16c$ and $16d$ Wyckoff sites; these mixed-occupancy columns are indistinguishable from the DRX phase. Higher-order axes, such as [211], are impractical due to the extreme alignment precision required for atomic resolution. In contrast, the [110] projection (Figure \ref{fgr:Structure}a--c) resolves ``pure $16c$" and ``pure $16d$" columns, enabling clear differentiation between the emerging spinel and parent DRX orders. However, the [110] projection cannot distinguish between different spinel variants. As demonstrated in Figures \ref{fig:100_Interface_Example}c and \ref{fig:cases}a--\ref{fig:cases}d, distinct variants produce identical projected orderings in simulated STEM-HAADF contrast. This reflects a general consequence of symmetry reduction on a parent lattice, in which distinct variants related by lost symmetry operations can share identical projections along specific zone axes. As a result, a spinel variant cannot be objectively identified from the ordering of a domain alone, but can only be described by its atomic displacement relative to other variants. The characterization of spinel domains must therefore be centered around that of interfaces, and use Fourier filtering to detect ordering changes which would not otherwise be obvious.

In a perfectly ordered single spinel variant, the two superlattice frequencies are fully phase-aligned, resulting in continuous $(\bar{1}1\bar{1})$ and $(\bar{1}11)$ fringes across the domain. This continuity is observed experimentally in the dashed circles of Figures \ref{fig:experimental}c and \ref{fig:experimental}d, and simulated in Figure \ref{fig:cases}d. This ideal case serves as a reference: any loss of fringe continuity indicates structural perturbations in $(\bar{1}1\bar{1})$ or $(\bar{1}11)$ plane periodicities. Depending on the relative displacement of the two variants, an interface may disrupt one, both, or neither of the two spinel plane spacings in projection. Table \ref{tab:discontinuity} sums up what Fourier filtering can theoretically reveal about \{100\} interfaces between all spinel variants of $\delta$-DRX, from the [110] zone axis. Fourier filtering enables the detection of 24 out of 28 simulated interfaces, and allows their classification among the four Fourier families described in Figures \ref{fig:cases}a, \ref{fig:cases}b, \ref{fig:cases}c, and \ref{fig:cases}d.

Table \ref{tab:discontinuity} displays trends creating a clear link between variant symmetry relationships, and the type of Fourier profile projected. Variant combinations resulting in no discontinuity are either both $A_i$, both $B_i$, both $C_{i}$, or both $D_i$. As seen in Figure \ref{fgr:Structure}b, each variant pair ($i=1$ and $i=2$) is related by a 2-fold axis of rotation, conserving the placement of pure $16d$ columns, pure $16c$ columns, and mixed columns.

Spinel variant pairings which create a discontinuity in both $(\bar{1}1\bar{1})$ and $(\bar{1}11)$ type fringes are either $A_i$--$B_i$ or $C_{i}$--$D_i$ combinations. These variant pairings, found in Figure \ref{fgr:Structure}b, are related to each other by complete exchange of their vacant and occupied sub-lattices ($16c$ and $16d$ sites). This sub-lattice exchange results in purely occupied columns becoming purely vacant, purely vacant columns becoming purely occupied, and mixed occupancy columns remaining mixed.

The simulated superposition of any such variant pair gives rise to a ``DRX-like" projection in the interface region, as it overlays the purely occupied columns of one variant with the purely vacant columns of the other, and vice versa. This results in the apparent filling of the vacant sites characteristic of ordered spinel in [110], which are needed to differentiate spinel from DRX. As shown in Figure \ref{fig:cases}c, the offset overlap of the two spinel orderings shortens the apparent spacing between equivalent columns in [110], creating a ``DRX-like" projected periodicity. This is seen in the STEM-HAADF simulations of Supplementary Figures S4, S5, S12, S13, S17, S18, S23 and S26, where each interfacial variant pair is likewise related. The triangular region in Figure~\ref{fig:experimental}a, where the projected contrast resembles that of the parent DRX lattice, shows an experimental situation where a DRX-like projection could potentially result from similar spinel variant overlap.

In the Fourier transform, a DRX-like spatial frequency indexed at the parent-lattice $(\bar{2}2\bar{2})/(\bar{2}22)$ positions is present as a consequence of the transformation from DRX, and is shown in the FFT of Figure~\ref{fig:cases}c, circled in white. In the Fourier-filtered images, this frequency appears as white ``DRX-like'' fringes because DRX-like frequencies are not suppressed in regions where the spinel fringes are discontinuous. When combined with the DRX-like appearance of the projected interfacial ordering, this can mimic remnant DRX in a multi-domain but otherwise highly ordered material.

Previous work on $\delta$-DRX has reported apparent excess remnant disorder in STEM-HAADF images filtered for spinel and DRX-like frequencies, in contrast with the much lower disorder estimated using SEND and ex-situ XRD \cite{holstun2025accelerating_Tucker, hau2024earth-abundant}. Our findings suggest that fringes seemingly belonging to DRX may result from frequencies generated by overlapping ordered spinel domains within the projection depth. This provides a plausible explanation for how projection overlap could inflate apparent disorder in filtered experimental STEM-HAADF where no real DRX is left in the transformed stucture, consistent with minimal remnant disorder in $\delta$-DRX.

In contrast to the complete sublattice exchange cases described above, other variant relationships produce only partial discontinuities and distinct projected morphologies. Spinel variant pairings which result in a discontinuity only in the $(\bar{1}1\bar{1})$ type fringes are either $A_i$--$C_{i}$ or $B_i$--$D_i$, which are related to each other by a translation in which $16c$-$16d$ mixed columns become purely $16c$ or purely $16d$, and where purely occupied and purely vacant columns become mixed.
Similarly, spinel variant pairings which result in a discontinuity only in the $(\bar{1}11)$ type fringes are either $A_i$--$D_i$ or $B_i$--$C_{i}$, which are also related by the change of purely occupied and purely vacant columns into mixed columns.

The projected interface of these variant pairs in Figures~\ref{fig:cases}a and \ref{fig:cases}b appear ``layered-like". In these thin-slab simulations, most of the intensity concentrates into alternating rows, giving the projection a layered-like appearance, while some very weak atomic column intensities remain visible between the rows. As the two variants overlap, the superposition of purely occupied and purely vacant columns from one variant with mixed columns to another produces alternating regions of increased and decreased simulated STEM-HAADF contrast. This is observed in the simulated STEM-HAADF contrast of Supplementary Figures S2-S29. A comparable type of contrast is seen in the experimental STEM-HAADF image (Figure~\ref{fig:experimental}a, marked as pentagon and square), where spots of much lower intensity are visible between rows of stronger contrast.

In filtered images, these interfaces produce DRX-like fringes confined to the disrupted spinel frequency. This type of projection overlap can therefore also contribute to the apparent surplus of disorder in filtered images. Our results suggest that largely ordered spinel domains can also produce layered-like contrast in projection, although the intensity of the weak inter-row spots observed experimentally cannot be anticipated from the simulated contrast alone. If layered-like features are observed in fully transformed $\delta$-DRX or related spinel-like Mn-rich cathodes, projection overlap between spinel domains provides a likely explanation.

Nevertheless, Table \ref{tab:discontinuity} does reveal a lack of structural specificity in simulated STEM-HAADF contrast, since many distinct interfaces share identical Fourier-filtered profiles. Even for detectable interfaces, exact variants are not identifiable. This imposes a fundamental limit on the ability of experimental STEM-HAADF imaging to confirm the presence of all eight predicted variants in the material. Some variant pairings in the simulations, such as $A_{1}$--$A_{2}$, interrupt neither $(\bar{1}1\bar{1})$ nor $(\bar{1}11)$ type fringes, rendering the boundary invisible altogether from the [110] zone axis.

These ambiguities arise directly from projection geometry. A displacement, or offset between variants has direction-dependent consequences: if its component normal to one plane family alters the projected interplanar spacing, only the associated spatial frequency will exhibit a discontinuity, while the other may remain unaffected. When the displacement contains components normal to both plane families, both frequencies are perturbed. Conversely, displacements that preserve the projected spacing within each plane family, or occur predominantly along the beam direction, may leave the Fourier profile unchanged despite underlying three-dimensional rearrangements. The projected morphology of both simulated interfaces and experimental antiphase boundaries can thus obscure the true structural relationship between variants.

Furthermore, the identification of ``DRX-like" or ``layered-like" interfacial projections may not be so straightforward. In experimental particles, multiple domains may intersect within the probed thickness, and real antiphase boundaries may curve or rotate into planes other than \{100\}. As a result, several boundaries can overlap along the beam direction, compressing the projected width and producing interfaces that appear much thinner in STEM-HAADF images than in our idealized slab simulations. The comparatively wide interfaces observed in the simulations arise from the specific geometries modeled here and should not be interpreted as quantitative predictions of antiphase boundary widths in physical $\delta$-DRX particles.

In addition, the results summarized in Table \ref{tab:discontinuity} are, in principle, applicable to interfaces beyond the  \{100\} family, since the Fourier discontinuity arises from the symmetry relationship between variants rather than from the specific plane along which the boundary forms. However, if an interface lies perpendicular to the beam direction, the projected ordering will include complete overlap of both domains without a detectable lateral shift, rendering the boundary invisible in projection regardless of its crystallographic character. 

Beyond these projection-specific constraints, our results carry broader implications for how local rearrangements and antiphase boundaries are interpreted in transformed materials, especially those with complex or slanted interface geometries. Crystallographically well-matched boundaries, like our low-energy \{100\} interfaces, may remain hidden in atomic-resolution STEM-HAADF, not because they lack structural complexity, but because projection superposition masks interfacial displacements. This has important implications for structure–property analysis, as materials accommodating discontinuity without large projected displacements may be systematically underrepresented in single-axis imaging.

Our results for $\delta$-DRX will therefore be of relevance to other systems that undergo local atomic reordering on a coherent lattice. For example, in Li--Ni--Mn--O spinel cathodes, operando multi-axis STEM and spectroscopy reveal antiphase boundaries and spatially inhomogeneous phase transformations during delithiation \cite{gong2018three}. These boundaries mediate local rearrangements while maintaining aspects of the spinel framework, and their projected appearance varies strongly with zone axis. This highlights orientation-dependent challenges in boundary visualization and interpretation, conceptually adjacent to those encountered in $\delta$-DRX.

Additional examples can be found in materials formed through local rearrangements, where domains and boundaries promote functional pathways. In nickelates for oxide electronics and superconductivity, the perovskite-to-infinite-layer topotactic reduction preserves the Ni–O network while selectively removing apical oxygen, producing locally ordered infinite-layer regions embedded within a perovskite host \cite{puphal2021topotactic}. This transformation involves substantial changes in coordination while preserving the parent lattice, meaning that distinct local structural states can coexist within a shared crystallographic framework. Such framework continuity can reduce structural contrast in projection, even though the underlying coordination environments differ. Furthermore, topotactically transformable antiphase boundaries have been engineered in oxide-ion and mixed-conductor materials, where interfaces that preserve the anion sublattice create connected pathways that enhance ionic transport \cite{xu2023topotactically}. A related framework-preserving mechanism governs the multi-electron conversion of magnetite, in which retention of the anion sublattice enables coherent phase evolution and stabilizes cycling performance \cite{zhang2019multi}. In these systems, the structural coherence that enables functional pathways may also reduce interpretability in projection, suggesting that the challenges identified in $\delta$-DRX are unlikely to be isolated and warrant careful scrutiny of locally generated domain structures in transformed materials. The examples discussed above illustrate that projection-induced ambiguities may arise as a general consequence of framework-preserving transformations in various multi-domain systems. In such cases, system-specific crystallographic analyses are required to identify hidden interfaces and avoid overinterpreting projection data. Once the crystallographically permitted domain relationships and their projected signatures are established, experimental images can be interpreted with substantially greater confidence. However, several experimental limitations must be carefully considered. For instance, increased dynamical diffraction in thicker specimens and experimental artifacts such as minor sample mistilt away from the exact favorable zone axis or scanning drift can degrade sublattice contrast or introduce artificial intensity asymmetries that falsely mimic atomic displacements, structural boundaries or strain \cite{sha2022deep_subangstrom, smeaton2023influence_light_atoms, burger2020influence_of_lens_aberrations} Finally, evaluating the global degree of disorder at the particle scale remains crucial. Since high-resolution STEM-HAADF features an inherently limited field of view, localized regions may appear perfectly ordered despite macroscopic heterogeneity \cite{hau2024earth-abundant, koh2025observation_order_disorder}. To mitigate this, larger-scale characterization techniques, such as scanning electron nano-diffraction or four-dimensional STEM (4D-STEM), are highly valuable for establishing the true degree of structural disorder prior to detailed atomic-level analysis.

On the characterization side, particularly in the context of domain characterization and antiphase boundary identification, mitigating these projection-induced ambiguities may require combining predictive modeling of transformation pathways with three-dimensional or diffraction-based methods that are sensitive to features hidden in projection. Complementary approaches such as multislice ptychography \cite{maiden2012ptychographic, romanov2024multi, li2018multi} or atomic-resolution tomography \cite{wang2020three, miao2016atomic} may mitigate projection artifacts and provide improved three-dimensional structural insight. In the case of the nickelates discussed above, multislice ptychography has recently allowed depth-resolved imaging of residual apical oxygen that is not reliably captured in conventional projected reconstructions, illustrating how phase-sensitive approaches can recover local structural information obscured in projection \cite{nickelate_ptychography_direct}. Multislice ptychography has also been demonstrated to recover the three-dimensional structure of buried heterointerfaces from a single-view dataset, providing depth sensitivity beyond the aperture-limited resolution of incoherent STEM imaging modes \cite{buried_interfaces_ptychography}.

In cases where domains or interfaces are expected to overlap in projection, STEM-HAADF can benefit from acquisition along multiple zone axes whenever feasible. Where overlap is expected, choosing a zone axis that separates parent and superlattice spacings can reduce ambiguity, and supplementing projection images with measurements not limited to column intensities (for example, diffraction or spectroscopy) can help resolve unclear features. In particular, techniques such as SEND provide reciprocal-space information at nanometer resolution and have been used to resolve local short-range ordering and symmetry changes that are difficult to access with conventional real-space imaging alone \cite{hsiao2022data, meng2016three, zeltmann2020patterned}. Likewise, X-ray diffraction offers a more robust measure of the overall extent of disorder \cite{Lun2020_Chem_Mn_DRX, wardini2021probing,li2024structural}, while STEM-HAADF remains useful for directly visualizing the size and morphology of domains. Together, these approaches may reduce the risk of mischaracterizing interface boundaries while keeping projection imaging useful.

\section*{Conclusions}
We show on this paper how antiphase boundaries between the eight distinct spinel variants that form from a rocksalt structure can present on atomically resolved STEM-HAADF. We find that some boundaries may remain undetected in projection, leading to potential errors in determining the distribution and morphology of domains. Even when boundaries are visible, multiple variant combinations can produce indistinguishable Fourier filtered profiles, limiting our ability to assign specific variant identities from single zone axis images alone.
Our results highlight that projection effects and interface overlap can exaggerate apparent disorder or generate projected motifs that do not seem to correspond to true spinel structures when relying solely on STEM-HAADF and Fourier-filtered images. These limitations underscore the need for careful interpretation and for integrating complementary techniques capable of probing antiphase boundaries beyond what is accessible from projected intensities.
The example used in this work, $\delta$-DRX,  belongs to a class of materials in which local rearrangements produce multiple crystallographic variants separated by antiphase boundaries. As such systems become increasingly relevant in energy storage and other functional applications, developing methods that provide reliable crystallographic information across length scales and zone axes will be essential. Such advances will enable more accurate structural analysis and a more complete understanding of how domain morphologies and antiphase boundaries relate to material behavior.

\section{Methods}
\subsection{Synthesis of $\delta$-DRX cathodes}
The DRX powder was synthesized using a molten-salt synthesis method. \ce{Li2CO3} (Sigma Aldrich, 99.9$\%$), \ce{Mn2O3} (Alfa Aesar, 99$\%$), \ce{TiO2} (Sigma Aldrich, 99.9$\%$), and LiF (Alfa Aesar, 99.9$\%$) were used as precursors. All precursors were mixed in stoichiometric proportions with ethanol in a Retsch PM 200 planetary ball mill at a rate of 250 rpm for 12 h. A 10$\%$ excess of \ce{Li2CO3} was added to compensate for possible loss during synthesis at 1100$^{\circ}$C. The precursors were dried in an oven at 70$^{\circ}$C overnight. The dried precursor was then mixed with KCl (Sigma Aldrich, 99$\%$) by hand for 10 minutes in a 1:3 precursor:KCl weight ratio before being pelletized. The pressed pellets were heated to 1100$^{\circ}$C at a rate of 10$^{\circ}$C min$^{-1}$ under argon flow in a tube furnace and held at this temperature for 20 minutes before naturally cooling to room temperature. For the molten-salt material, the heated pellet was washed directly with deionized water twice and ethanol once, followed by drying at 70$^{\circ}$C overnight under vacuum. The obtained powder was then transferred to and stored in an argon-filled glovebox.

Cathode films were composed of active material, Super C65 (Timcal), and polytetrafluoroethylene (PTFE, DuPont, Teflon 8A) in a weight ratio of 70:20:10. Prior to cathode fabrication, the active material was mixed with Super C65 by hand in a mortar and pestle for 20 minutes. This mixture was then manually mixed with PTFE in a mortar and pestle for 10 minutes before being rolled into thin films inside a glovebox. The active material loading density of all films was $\approx$3–4 mg cm$^{-2}$. Commercial 1 M \ce{LiPF6} in ethylene carbonate (EC) and dimethyl carbonate (DMC) solution (1:1 volume ratio) was used as the electrolyte. A glass microfiber filter (Whatman) was used as the separator. FMC Li metal foil was used as the anode. Coin cells were assembled inside the glovebox.

An electrochemical pulsing procedure was conducted to transform these DRX electrode films to the $\delta$-DRX phase. Cells were cycled in multi-zone temperature chambers set to 50$^{\circ}$C and connected to an Arbin test instrument. As described in our previous publication\cite{holstun2025accelerating_Tucker}, the electrochemical pulsing consisted of an initial constant current step at $\pm$10 A g$^{-1}$ until a cutoff of either 4.5 V or 3.0 V was reached, followed by a constant voltage hold until the current dropped to $\pm$10 mA g$^{-1}$. After pulsing, cells were charged to 3.5 V and held for 6 hours prior to disassembly for TEM.

\subsection{TEM lamella preparation and atomic resolved STEM-HAADF imaging}
The ex situ samples for atomic-resolution TEM imaging were prepared by mixing the $\delta$-DRX cathode particles with PTFE in a 9:1 ratio using a mortar and pestle. This composite of the cathode active material and PTFE was then rolled into a cathode thin film. A thin lamella suitable for TEM analysis was prepared from the cathode thin film using a Helios G4 UX dual-beam focused ion beam (FIB) instrument.The lamella was thinned down to $\sim$1 $\mu$m using a gallium-ion beam with a 30 kV accelerating voltage and a 0.75 nA current. To prevent ion beam damage, subsequent thinning to $\sim$500 nm was performed using an 8 kV accelerating voltage and a 0.15 nA current, followed by thinning to $\sim$100 nm thickness using a 5 kV accelerating voltage. To remove the amorphous layer and further thin the sample, the voltage was dropped to a 2 kV beam, followed by final polishing at 1 kV.

The atomic resolution STEM-HAADF imaging was performed using the TEAM I microscope (double aberration- corrected Thermo Fisher Scientific Titan 80–300 kV) housed at NCEM, Berkeley. The microscope was operated at 300 kV with an indicated convergence angle of 30 mrad. The measurements were carried out with a total electron dose of $\approx$ 4 $\times$ $10^4$ $e^-/\AA^2$ was used. The particle was tilted to [110] zone axis before atomic resolved imaging to differentiate between the spinel and the DRX phases. The final micrographs were obtained after correcting for the non-linear drift distortion by using image pairs with orthogonal scan directions using the quantEM package \cite{ophus2016correcting}. 

\subsection{Structural model generation of the eight crystallographic variants}

All crystallographic variant models were generated using Python scripts based on Pymatgen. The starting structure corresponds to a primitive rocksalt lattice ($a=b=c=2.96985$~\AA, $\alpha=\beta=\gamma=60^\circ$, space group $P1$) containing a partially occupied, cation disordered transition-metal/Li octahedral sublattice. This disordered parent structure was expanded into oriented supercells using integer transformation matrices generated.

To define ordered spinel-like octahedral occupancy patterns within these supercells, a reference ordered spinel structure, Li$_2$Mn$_4$O$_8$ was separately expanded and rigidly registered to the disordered supercell by translation and rotation operations. Four unique rotational variants were generated by aligning a tetrahedral cluster of reference Mn sites to a corresponding cluster in the target supercell, thereby defining four symmetry-related Mn occupation sets on the octahedral sublattice. Octahedral sites were assigned as occupied if their Cartesian coordinates matched a mapped reference Mn position within a tolerance of 0.2~\AA. The remaining four variants were generated as translational complements of the corresponding rotational variants by inverting octahedral occupancy (occupied $\leftrightarrow$ vacant) across the full octahedral sublattice. This procedure yielded eight distinct crystallographic variants, stored as octahedral-site occupancy index lists for subsequent interface construction.

For visualization of the minimal repeating unit in Fig.~\ref{fgr:Structure}b--c, reduced [110] projections of the variant models were rendered using VESTA. Vacant sites were optionally filled with placeholder atoms solely for schematic visualization.

\subsection{Simulation of a two-variant slanted interface structure}

Two-variant interface supercells were generated using a Python workflow based on Pymatgen and Numpy.
Interface slabs were constructed with a [110] slab orientation such that octahedral-site ordering remains distinguishable in projection along [110]. A planar domain boundary corresponding to a $\{100\}$ interface was introduced by defining a slanted partition in fractional coordinates using a linear function of the form
\begin{equation}
f(y,z) = z - m y - b,
\end{equation}
where $m$ controls the interface tilt and $b$ sets the interface position within the supercell. Each lattice site was assigned to one of two crystallographic variants based on the sign of $f(y,z)$ evaluated at its fractional coordinates: sites satisfying $f(y,z) < 0$ were populated according to one variant ordering, while sites satisfying $f(y,z) > 0$ were populated according to the other variant ordering. The parameters $m$ and $b$ were chosen such that the interface lies near the center of the supercell. This geometry produces a $\{100\}$ interface which intersects [110] at a 45° angle.

The supercell containing the slanted interface structure had dimensions $a=8.07173$~\AA, $b=79.90604$~\AA, and $c=34.24545$~\AA. To increase the simulated slab dimensions without regenerating the interface geometry, was subsequentely tiled 15 times along the $a$  [001] direction prior to HAADF simulation. The final simulation supercell had dimensions $a=121.07593$~\AA, $b=79.90604$~\AA, and $c=34.24545$~\AA~ and contained 30240 atoms (\ce{Mn10080O20160}). 
The simulated slab thickness used for the interface analysis was 3.42 nm. Although the experimental STEM-HAADF images were acquired from lamellae approximately 100 nm thick, the experiments were performed using a convergence semi-angle of approximately 30 mrad, corresponding to an effective depth of field of approximately 3.5 nm under these imaging conditions.\cite{lupini2011_aberration_depth} This depth of field is comparable to the simulated slab thickness, supporting the comparison between the simulated and experimental images. This interface creation and supercell expansion procedure was repeated for all variant pair combinations (variants 0 to 7). Final interface geometries were visually inspected in VESTA to verify the intended domain boundary placement and orientation.

\subsection{Simulation of HAADF micrographs using projected potential}

The simulated HAADF micrographs were generated from the projected potentials of the interface supercells using the abTEM Python package\cite{madsen2021abtem},  rather than by performing a full multislice STEM image simulation including explicit electron-probe propagation. Atomic structures were imported as VASP-format files using the Atomic Simulation Environment (ASE). To isolate the contribution of the transition-metal sublattice responsible for the spinel cation ordering, only Mn atoms were retained for subsequent simulation steps, and all non-Mn species were removed from the atomic model prior to potential construction.

The Mn electrostatic potential was computed using abTEM  with a real-space sampling of 0.2~\AA\ per pixel, an infinite projection approximation, a slice thickness of 1~\AA, and the Lobato parametrization of electron scattering factors\cite{lobato2014accurate}.The electrostatic potential of the Mn sublattice was calculated from the atomic structure and projected along the [110] direction (c-axis of the supercell containing the interface), yielding a two-dimensional projected potential map.

To generate an intensity map representative of Z-contrast imaging suitable for Fourier analysis, the projected potential was squared. The squared-potential image was subsequently Gaussian blurred ($\sigma$ = 3 pixels) and normalized to the range [0,1] prior to export. The simulation images were exported as both NumPy arrays and as rendered PNG files for visualization.

\subsection{Fourier filtering and characterization of spinel variant boundaries}

Fourier filtering was applied to both experimental STEM-HAADF images and simulated HAADF-like projected potential maps to isolate spinel superlattice periodicities in the [110] projection and to evaluate domain boundary discontinuities.

For both experimental and simulated datasets, the resulting micrographs were Fourier transformed using a 2D FFT, and the complex fourier spectrum was analyzed to identify frequency components corresponding to parent DRX lattice periodicities and spinel superlattice periodicities. Circular reciprocal-space masks were constructed around the two non-equivalent spinel superlattice reflections observable in the [110] projection, and inverse Fourier transforms were used to generate real-space fringe images corresponding to the selected periodicities.

To minimize spectral leakage and edge artifacts, images were apodized prior to Fourier transformation using a 2D Tukey (tapered cosine) window. Fourier coefficients outside the masked regions were attenuated using a soft mask to avoid sharp cutoff artifacts. Additional masks were applied at the expected DRX peak positions to suppress contributions from the parent lattice and facilitate visualization of spinel fringe continuity.

Filtering parameters (including mask radii and Fourier peak positions in pixel coordinates) were adjusted slightly between experimental and simulated datasets to account for differences in image sampling and field of view. Supplementary figure S31 illustrates the effect of varying the mask radius size by comparing the filtered results using two different mask radii, for the same interface.

Domain boundaries were identified by comparing the continuity of the filtered fringe contrast across the interface region in each reconstructed component; interfaces were classified as continuous or discontinuous depending on whether the filtered fringe phase and intensity remained coherent across the boundary or exhibited a detectable shift or interruption.

\section*{Author contributions}
\textbf{Ninon Scherz:} Methodology, Software, Validation, Formal analysis, Investigation, Data Curation, Writing-Original Draft, and Visualization. 
\textbf{Shashwat Anand:} Investigation, Methodology, Software, Writing-Original Draft, and formal analysis. 
\textbf{Colin Ophus:} Writing-review \& editing and methodology.
\textbf{Tucker Holstun:} Investigation and Writing-review \& editing. 
\textbf{Mary Scott:} Writing-review \& editing. 
\textbf{Tara P. Mishra:} Conceptualization, Methodology, Validation, Investigation, Writing-Original Draft, Visualization, Supervision. 
\textbf{Gerbrand Ceder:} Writing-review \& editing, Supervision, Resources, Funding acquistion. 

\section*{Conflicts of interest}
There are no conflicts to declare

\section*{Data availability}

The data supporting this article have been included as part of the Supplementary Information.

\section*{Acknowledgements}

This work was supported by the Assistant Secretary for Critical Minerals and Energy Innovation, Transportation Technologies Office, under the Applied Battery Materials Program, of the U.S. Department of Energy under Contract No. DE-AC02-05CH11231. Work at the Molecular Foundry was supported by the Office of Science, Office of Basic Energy Sciences, of the U.S. Department of Energy under Contract No. DE-AC02-05CH11231, under proposal number(s) MFP-8834 and MFP-8187. CO acknowledges funding from the the US DOE Early Career Research Program



\balance


\section{References}
\bibliography{rsc_delta_ref} 
\bibliographystyle{rsc} 
\end{document}